\documentclass[10pt,letterpaper]{article}
\usepackage[top=0.85in,left=2.75in,footskip=0.75in]{geometry}

\usepackage{amsmath,amssymb}

\usepackage{changepage}

\usepackage[utf8x]{inputenc}

\usepackage{textcomp,marvosym}

\usepackage{cite}

\usepackage{nameref,hyperref}

\usepackage[right]{lineno}

\usepackage{microtype}
\DisableLigatures[f]{encoding = *, family = * }

\usepackage[table]{xcolor}

\usepackage{array}

\usepackage{rotating,tabularx}

\newcolumntype{+}{!{\vrule width 2pt}}

\newlength\savedwidth

\raggedright
\usepackage[aboveskip=1pt,labelfont=bf,labelsep=period,justification=raggedright,singlelinecheck=off]{caption}

\makeatletter
\renewcommand{\@biblabel}[1]{\quad#1.}
\makeatother

\date{}

\usepackage{lastpage,fancyhdr,graphicx}
\usepackage{epstopdf}
\fancyheadoffset[L]{2.25in}
\fancyfootoffset[L]{2.25in}

\begin{document}

\vspace*{0.2in}

\begin{flushleft}
{\Large
\textbf\newline{The quantitative measure and statistical distribution of fame} 
}
\newline
\\
Edward D. Ramirez and
Stephen J. Hagen*

\bigskip
Physics Department, University of Florida, Gainesville Florida 32611-8440 USA
\\

\bigskip

%
%





* sjhagen@ufl.edu

\end{flushleft}
\section*{Abstract}
Fame and celebrity play an ever-increasing role in our culture.  However, despite the cultural and economic importance of fame and its gradations, there exists no consensus method for quantifying the fame of an individual, or of comparing that of two individuals. We argue that, even if fame is difficult to measure with precision, one may develop useful metrics for fame that correlate well with intuition and that remain reasonably stable over time. Using datasets of recently deceased individuals who were highly renowned, we have evaluated several internet-based methods for quantifying fame.  We find that some widely-used internet-derived metrics, such as search engine results, correlate poorly with human subject judgments of fame. However other metrics exist that agree well with human judgments and appear to offer workable, easily accessible measures of fame. Using such a metric we perform a preliminary investigation of the statistical distribution of fame, which has some of the power law character seen in other natural and social phenomena such as landslides and market crashes.  In order to demonstrate how such findings can generate quantitative insight into celebrity culture, we assess some folk ideas regarding the frequency distribution and  apparent clustering of celebrity deaths.



\section*{Introduction}

The phenomena of fame and celebrity are increasingly important in our culture. With the rapid expansion of electronic media, fame plays a growing role in commerce, media, and public affairs, as well as in legal and academic spheres \cite{RefWorks:911}. Social media have boosted the visibility of celebrities of all kinds, allowing individuals to acquire or lose fame overnight \cite{RefWorks:903}.   Celebrity endorsements offer value to businesses, political campaigns and cultural organizations. Fame affects the economic value of names and trademarks \cite{RefWorks:948}, and it aids professional advancement in a variety of fields. 

Researchers have explored some aspects of fame, such as the psychological motives of the would-be famous \cite{RefWorks:952} and sex bias in assessments of fame  \cite{RefWorks:904}. Several studies have also attempted to correlate the fame of well known individuals with measures of their professional achievement \cite{RefWorks:859} \cite{RefWorks:922}  \cite{RefWorks:923}  \cite{RefWorks:906} \cite{RefWorks:860} \cite{RefWorks:949},  \textcolor{black}{which is measured by different tools in different fields.} Remarkably however, although fame clearly exists in degrees there is no consensus on its quantitative measure: researchers who have attempted to quantify fame have relied on a variety of ad hoc measures \textcolor{black}{ that have not themselves been evaluated or calibrated.} 

One common such measure has been search engine results.  Following Schulman's proposal \cite{RefWorks:957} that an individual's fame is revealed by the number of web pages returned from an internet search for his/her name, many researchers have used Google hits to quantify fame. Google hits (denoted \emph{GH}, the number of web pages returned in a Google search for an individual's name) has been used to quantify the fame of WWI flying aces \cite{RefWorks:859} and chess masters \cite{RefWorks:906} as well as physicists \cite{RefWorks:860}. Similarly, some researchers have used Wikipedia data, including page views and other measures of Wikipedia presence, to quantify the fame of athletes \cite{RefWorks:923} and historical figures \cite{RefWorks:949} and to predict movie box office success \cite{RefWorks:944}. \textcolor{black}{Other social media tools have also been employed to measure attributes related to fame; one study used the number of Twitter followers \cite{RefWorks:922} to gauge the social media visibility of a sampling of scientists.} 
	
There are alternatives to using internet tools or social media to measure fame. Psychological researchers have studied how panels of human subjects judge the fame of well-known individuals \cite{RefWorks:904}  \cite{RefWorks:950}.  A recent  ``culturomics'' study tracked the rise and fall of individuals' fame on historical time scales by measuring the frequency of their mention in a large database of digitized texts \cite{RefWorks:903}. Some journalists regard the length of an individual's obituary, as well as its advance preparation, as indices of fame \cite{RefWorks:953}.    

Nevertheless internet tools such as \emph{GH} are convenient to use. Unfortunately, researchers have generally not attempted to validate these tools by testing whether they give results consistent with other measures or intuitive indicators of fame.  It is remarkable that no metric has been tested for measuring a phenomenon that is unequally distributed and yet has demonstrable utility and economic value across fields.  Furthermore, \textcolor{black}{studies of fame have often failed to define it separately from related  concepts such as celebrity, professional accomplishment, and media profile.}  This lack of precision hinders the quantitative study of fame and its accurate valuation. It also prevents serious assessment of common claims about celebrity and fame: The often-made assertions that an unusual number of famous individuals died within a given year \cite{RefWorks:954}, or that famous deaths occur in clusters \cite{RefWorks:955}, or that famous musicians die young \cite{RefWorks:977}, cannot be assessed unless fame can be quantified. \textcolor{black}{This article aims to demonstrate that, starting from a clear definition of fame, internet-derived metrics of fame can be found that correlate with quantitative human judgments of fame,} and that using these metrics one may gain insight into some of the statistical properties of fame.  

\textcolor{black}{For clarity we begin by defining an individual’s fame as his/her degree of renown, or state of being well known, to a population. We do not assume that the fame of an individual correlates with accomplishment as judged by that population, or that the metrics of accomplishment favored by that population are also metrics of fame.}

\textcolor{black}{ In contrast we define celebrity as the close media attention that is provided to the most famous individuals; thus a celebrity is one whose ordinary activities receive media attention.  Fame and celebrity correlate, but they are not the same.}  These definitions accord with those of Drake and Miah \cite{RefWorks:911}, who described a celebrity as a mediated public persona. 

By our definition the fame of an individual \textcolor{black}{can be measured as a snapshot taken from the perspective of a given population at a particular time.  Metrics that gauge renown among different populations at different times may not agree completely.  We do not attempt to measure the fame that individuals may have enjoyed in the past, or the peak fame that they achieved.  Rather our approach is to study a diverse group of renowned individuals at one common time point in their career – the year following their death – and to quantify their fame at that time point. We do this by measuring their renown among a group of survey subjects.  The survey data provides a baseline, quantitative fame score that we then compare against some plausible internet or social media metrics of fame.   }   In this way we identify metrics of fame that can be easily employed on a larger scale to evaluate the renown of many individuals.  \textcolor{black}{We then use one such metric to investigate some statistical properties of fame and demonstrate how these statistics can provide insight into some folk ideas regarding the frequency of famous deaths.} 

\section*{Methods}

\subsection*{Data sources}

\textcolor{black}{We investigated the fame of deceased individuals only. This is in part because we intended to use the findings to test folk claims regarding the frequency of celebrity deaths. However we also sought to minimize concerns related to name ownership and consent among the individuals whose fame we were evaluating. We also limited our study to those who had died within the year or so prior to this investigation, so as to avoid having  to devise corrections for possible changes in the fame of individuals after their death.}

We generated three lists (denoted NBC, Wiki, and NYT) of renowned individuals who died in 2016 or 2017. No individual appeared in more than one list. The NBC list consisted of 126 highly renowned individuals whose deaths occurred during the full year 2016 and received mention in NBC Online \cite{RefWorks:956}.  The Wiki list consisted of 78 names drawn at random from 642 individuals who (as of March 2017) were named on Wikipedia.org as having died during the month of January 2017 \cite{RefWorks:946}. The NYT list consisted of two parts, totaling 147 individuals who were named in the New York Times online obituaries \cite{RefWorks:947} as having died during two months in 2017; One part (NYT 1) is 75 individuals who died in February 2017, while the other part (NYT 2) is 72 individuals who died in June 2017.

\subsection*{Survey metric ($p$ ratings) }

\textcolor{black}{In order to generate an intuitive and quantitative scale for fame, against which we could compare various other possible metrics for fame, we first used a survey, based on pairwise comparisons, to rank a list of twenty famous individuals according to their renown. The individuals named on the survey are listed in Table~\ref{table1}. They were selected on the basis of (1) having died during 2016, with their obituaries widely reported in news media, and therefore being plausibly described as famous; (2) spanning a sufficient range in renown that statistically significant differences in their rankings could emerge from the data analysis; (3) being known in fields for which the survey subjects likely  possessed relevant general knowledge. Fifty undergraduate students at the University of Florida were recruited as subjects to complete the survey. Therefore the list in Table~\ref{table1} is a sampling of major political and historical figures, top American athletes, stars of popular films and music, authors of books often read by students, and similar figures. The list excludes individuals associated with more specialized interests, such as cabinet secretaries, academics, playwrights, foreign athletes, classical musicians, and so forth. }   

\renewcommand{\arraystretch}{1.4}

\begin{sidewaystable}
\centering
\caption{
{\bf Fame metrics for twenty renowned individuals}}	
\begin{tabular}{llllllllllll}
\rowcolor[HTML]{C0C0C0} 
\textbf{\boldmath{ID}} &
\textbf{\boldmath{$Names$}} & \textbf{\boldmath{$DOB$}} & \textbf{\boldmath{$DOD$}} & \textbf{\boldmath{$p$}} & \textbf{\boldmath{$\delta p$}} & \textbf{\boldmath{$WE$}} & \textbf{\boldmath{$GN$}} & \textbf{\boldmath{$GH$}} & \textbf{\boldmath{$WV$}} & \textbf{$\frac{d WE}{d t}$} & \textbf{$\frac{d WV}{dt}$} \\ \hline
\multicolumn{1}{|l|}{01} & \multicolumn{1}{l|}{Muhammad Ali} & \multicolumn{1}{l|}{1/17/1942} & \multicolumn{1}{l|}{6/3/2016} & \multicolumn{1}{l|}{0.18} & \multicolumn{1}{l|}{0.03} & \multicolumn{1}{l|}{10,909} & \multicolumn{1}{l|}{280,000} & \multicolumn{1}{l|}{69,300,000} & \multicolumn{1}{l|}{250,833} & \multicolumn{1}{l|}{61} & \multicolumn{1}{l|}{36,516} \\ \hline
\multicolumn{1}{|l|}{02} & \multicolumn{1}{l|}{Fidel Castro} & \multicolumn{1}{l|}{8/13/1926} & \multicolumn{1}{l|}{11/25/2016} & \multicolumn{1}{l|}{0.17} & \multicolumn{1}{l|}{0.03} & \multicolumn{1}{l|}{13,975} & \multicolumn{1}{l|}{149,000} & \multicolumn{1}{l|}{39,800,000} & \multicolumn{1}{l|}{110,573} & \multicolumn{1}{l|}{76} & \multicolumn{1}{l|}{13,685} \\ \hline
\multicolumn{1}{|l|}{03} & \multicolumn{1}{l|}{Prince} & \multicolumn{1}{l|}{6/7/1958} & \multicolumn{1}{l|}{4/21/2016} & \multicolumn{1}{l|}{0.17} & \multicolumn{1}{l|}{0.03} & \multicolumn{1}{l|}{10,102} & \multicolumn{1}{l|}{2,520,000} & \multicolumn{1}{l|}{768,000,000} & \multicolumn{1}{l|}{199,107} & \multicolumn{1}{l|}{56} & \multicolumn{1}{l|}{47,565} \\ \hline
\multicolumn{1}{|l|}{04} & \multicolumn{1}{l|}{Nancy Reagan} & \multicolumn{1}{l|}{7/6/1921} & \multicolumn{1}{l|}{3/6/2016} & \multicolumn{1}{l|}{0.079} & \multicolumn{1}{l|}{0.012} & \multicolumn{1}{l|}{3,664} & \multicolumn{1}{l|}{23,200} & \multicolumn{1}{l|}{13,200,000} & \multicolumn{1}{l|}{42,626} & \multicolumn{1}{l|}{19} & \multicolumn{1}{l|}{5,904} \\ \hline
\multicolumn{1}{|l|}{05} & \multicolumn{1}{l|}{Arnold Palmer} & \multicolumn{1}{l|}{9/10/1929} & \multicolumn{1}{l|}{9/25/2016} & \multicolumn{1}{l|}{0.062} & \multicolumn{1}{l|}{0.011} & \multicolumn{1}{l|}{1,933} & \multicolumn{1}{l|}{112,000} & \multicolumn{1}{l|}{36,300,000} & \multicolumn{1}{l|}{44,455} & \multicolumn{1}{l|}{11} & \multicolumn{1}{l|}{3,961} \\ \hline
\multicolumn{1}{|l|}{06} & \multicolumn{1}{l|}{Alan Rickman} & \multicolumn{1}{l|}{2/21/1946} & \multicolumn{1}{l|}{1/15/2016} & \multicolumn{1}{l|}{0.057} & \multicolumn{1}{l|}{0.010} & \multicolumn{1}{l|}{3,666} & \multicolumn{1}{l|}{62,000} & \multicolumn{1}{l|}{585,000} & \multicolumn{1}{l|}{117,690} & \multicolumn{1}{l|}{20} & \multicolumn{1}{l|}{19,510} \\ \hline
\multicolumn{1}{|l|}{07} & \multicolumn{1}{l|}{Harper Lee} & \multicolumn{1}{l|}{4/28/1926} & \multicolumn{1}{l|}{2/19/2016} & \multicolumn{1}{l|}{0.042} & \multicolumn{1}{l|}{0.007} & \multicolumn{1}{l|}{4,193} & \multicolumn{1}{l|}{15,100} & \multicolumn{1}{l|}{59,600,000} & \multicolumn{1}{l|}{32,566} & \multicolumn{1}{l|}{24} & \multicolumn{1}{l|}{5,384} \\ \hline
\multicolumn{1}{|l|}{08} & \multicolumn{1}{l|}{George Michael} & \multicolumn{1}{l|}{6/25/1963} & \multicolumn{1}{l|}{12/25/2016} & \multicolumn{1}{l|}{0.040} & \multicolumn{1}{l|}{0.007} & \multicolumn{1}{l|}{7,025} & \multicolumn{1}{l|}{432,000} & \multicolumn{1}{l|}{403,000,000} & \multicolumn{1}{l|}{115,813} & \multicolumn{1}{l|}{39} & \multicolumn{1}{l|}{3,652} \\ \hline
\multicolumn{1}{|l|}{09} & \multicolumn{1}{l|}{John Glenn} & \multicolumn{1}{l|}{7/18/1921} & \multicolumn{1}{l|}{12/8/2016} & \multicolumn{1}{l|}{0.037} & \multicolumn{1}{l|}{0.007} & \multicolumn{1}{l|}{3,920} & \multicolumn{1}{l|}{84,000} & \multicolumn{1}{l|}{125,000,000} & \multicolumn{1}{l|}{72,615} & \multicolumn{1}{l|}{23} & \multicolumn{1}{l|}{3,088} \\ \hline
\multicolumn{1}{|l|}{10} & \multicolumn{1}{l|}{Debbie Reynolds} & \multicolumn{1}{l|}{4/1/1932} & \multicolumn{1}{l|}{12/28/2016} & \multicolumn{1}{l|}{0.036} & \multicolumn{1}{l|}{0.006} & \multicolumn{1}{l|}{2,037} & \multicolumn{1}{l|}{95,900} & \multicolumn{1}{l|}{27,300,000} & \multicolumn{1}{l|}{118,990} & \multicolumn{1}{l|}{12} & \multicolumn{1}{l|}{3,934} \\ \hline
\multicolumn{1}{|l|}{11} & \multicolumn{1}{l|}{Gene Wilder} & \multicolumn{1}{l|}{6/11/1933} & \multicolumn{1}{l|}{8/29/2016} & \multicolumn{1}{l|}{0.032} & \multicolumn{1}{l|}{0.006} & \multicolumn{1}{l|}{2,303} & \multicolumn{1}{l|}{20,400} & \multicolumn{1}{l|}{10,600,000} & \multicolumn{1}{l|}{85,083} & \multicolumn{1}{l|}{14} & \multicolumn{1}{l|}{10,458} \\ \hline
\multicolumn{1}{|l|}{12} & \multicolumn{1}{l|}{Christina Grimmie} & \multicolumn{1}{l|}{3/12/1994} & \multicolumn{1}{l|}{6/10/2016} & \multicolumn{1}{l|}{0.018} & \multicolumn{1}{l|}{0.004} & \multicolumn{1}{l|}{3,071} & \multicolumn{1}{l|}{21,500} & \multicolumn{1}{l|}{631,000} & \multicolumn{1}{l|}{206,297} & \multicolumn{1}{l|}{42} & \multicolumn{1}{l|}{11,236} \\ \hline
\multicolumn{1}{|l|}{13} & \multicolumn{1}{l|}{Bill Paxton} & \multicolumn{1}{l|}{5/17/1955} & \multicolumn{1}{l|}{2/25/2017} & \multicolumn{1}{l|}{0.018} & \multicolumn{1}{l|}{0.004} & \multicolumn{1}{l|}{1,530} & \multicolumn{1}{l|}{1,220,000} & \multicolumn{1}{l|}{30,500,000} & \multicolumn{1}{l|}{166,144} & \multicolumn{1}{l|}{10} & \multicolumn{1}{l|}{2,580} \\ \hline
\multicolumn{1}{|l|}{14} & \multicolumn{1}{l|}{Kimbo Slice} & \multicolumn{1}{l|}{2/8/1974} & \multicolumn{1}{l|}{6/6/2016} & \multicolumn{1}{l|}{0.016} & \multicolumn{1}{l|}{0.003} & \multicolumn{1}{l|}{4,213} & \multicolumn{1}{l|}{13,000} & \multicolumn{1}{l|}{540,000} & \multicolumn{1}{l|}{109,950} & \multicolumn{1}{l|}{29} & \multicolumn{1}{l|}{8,307} \\ \hline
\multicolumn{1}{|l|}{15} & \multicolumn{1}{l|}{Elie Wiesel} & \multicolumn{1}{l|}{9/30/1928} & \multicolumn{1}{l|}{7/2/2016} & \multicolumn{1}{l|}{0.015} & \multicolumn{1}{l|}{0.003} & \multicolumn{1}{l|}{5,637} & \multicolumn{1}{l|}{8,960} & \multicolumn{1}{l|}{6,560,000} & \multicolumn{1}{l|}{33,537} & \multicolumn{1}{l|}{31} & \multicolumn{1}{l|}{2,700} \\ \hline
\multicolumn{1}{|l|}{16} & \multicolumn{1}{l|}{Juan Gabriel} & \multicolumn{1}{l|}{1/7/1950} & \multicolumn{1}{l|}{8/28/2016} & \multicolumn{1}{l|}{0.012} & \multicolumn{1}{l|}{0.003} & \multicolumn{1}{l|}{1,399} & \multicolumn{1}{l|}{41,500} & \multicolumn{1}{l|}{78,600,000} & \multicolumn{1}{l|}{11,594} & \multicolumn{1}{l|}{9} & \multicolumn{1}{l|}{3,375} \\ \hline
\multicolumn{1}{|l|}{17} & \multicolumn{1}{l|}{Vanity} & \multicolumn{1}{l|}{1/4/1959} & \multicolumn{1}{l|}{2/15/2016} & \multicolumn{1}{l|}{0.0081} & \multicolumn{1}{l|}{0.0020} & \multicolumn{1}{l|}{1,040} & \multicolumn{1}{l|}{1,480} & \multicolumn{1}{l|}{8,970,000} & \multicolumn{1}{l|}{17,233} & \multicolumn{1}{l|}{7} & \multicolumn{1}{l|}{3,771} \\ \hline
\multicolumn{1}{|l|}{18} & \multicolumn{1}{l|}{Keith Emerson} & \multicolumn{1}{l|}{11/2/1944} & \multicolumn{1}{l|}{3/11/2016} & \multicolumn{1}{l|}{0.0041} & \multicolumn{1}{l|}{0.0013} & \multicolumn{1}{l|}{1,629} & \multicolumn{1}{l|}{7,170} & \multicolumn{1}{l|}{4,370,000} & \multicolumn{1}{l|}{13,816} & \multicolumn{1}{l|}{9} & \multicolumn{1}{l|}{1,541} \\ \hline
\multicolumn{1}{|l|}{19} & \multicolumn{1}{l|}{Phife Dawg} & \multicolumn{1}{l|}{11/20/1970} & \multicolumn{1}{l|}{3/22/2016} & \multicolumn{1}{l|}{0.0038} & \multicolumn{1}{l|}{0.0012} & \multicolumn{1}{l|}{525} & \multicolumn{1}{l|}{6,700} & \multicolumn{1}{l|}{459,000} & \multicolumn{1}{l|}{16,783} & \multicolumn{1}{l|}{4} & \multicolumn{1}{l|}{2,579} \\ \hline
\multicolumn{1}{|l|}{20} & \multicolumn{1}{l|}{Afeni Shakur} & \multicolumn{1}{l|}{1/10/1947} & \multicolumn{1}{l|}{5/2/2016} & \multicolumn{1}{l|}{0.0029} & \multicolumn{1}{l|}{0.0010} & \multicolumn{1}{l|}{736} & \multicolumn{1}{l|}{3,220} & \multicolumn{1}{l|}{355,000} & \multicolumn{1}{l|}{128,852} & \multicolumn{1}{l|}{4} & \multicolumn{1}{l|}{1,585} \\ \hline
\end{tabular}
\label{table1}
\end{sidewaystable}

Each survey subject was presented with a list of fifty different pairs of names, drawn from the twenty names in Table~\ref{table1}. Each pair of names could be presented in either order ($A:B$ or $B:A$). The survey subject was asked to indicate a preference within each pair by identifying the name about which he/she had greater knowledge.  The subject could also select a ``no preference'' option if he/she felt equally knowledgeable about both names. The fifty pairings on each survey form were computer-selected at random from the 380 possible pairs that can be generated from twenty names. Each subject received a unique, randomized version of the survey. \textcolor{black}{The list of names was limited to twenty so that each of the possible name pairs  could be presented to multiple survey subjects, without requiring a survey of excessive length. Thus, with twenty names and fifty subjects, each being offered fifty comparisons, the survey offered each pair of names to approximately 13 subjects.  If instead 40 names had been tested, then 1560 name pairings would be possible and either the number of subjects or the length of the survey would have had to increase fourfold to achieve the same coverage.} 

\textcolor{black}{Of the fifty name pairs offered to each subject, subjects responded with an average (and median) of 34 preferences and 16 ``no preference'' responses. 86\% of subjects indicated a preference in at least half of the fifty pairs they were offered. Consequently, of the 2500 name pairs (50 subjects $\times$ 50 name pairs) offered to all subjects, 1679 elicited preferences and 821 elicited a ``no preference'' response from the subject. The preference data are provided in \nameref{S1_Spreadsheet}.}

\textcolor{black}{The ``no preference'' response could indicate that the subject was equally familiar with both names (two very famous names), or that the subject was equally unfamiliar with both names (two less famous names). Regardless of its cause, a ``no preference'' response does not facilitate the ranking of those two particular names by renown. As the purpose of the survey was to differentiate the individuals by renown, the ``no preference'' responses were omitted from the subsequent data analysis. The effect of survey sample size, including these omitted ``no preference'' responses, on the robustness of the obtained ranking was tested through (1) a bootstrap error analysis, discussed below, and (2) a log likelihood test, discussed in \textit{Survey Results}.}

We used a Bradley-Terry model \cite{RefWorks:912} to convert the preference data to a quantitative measure of fame, assigning a rating $p_i$ to each individual $i$ ($i = 1 \ldots 20$) in Table~\ref{table1}.  In the Bradley-Terry model, the strength scores or ratings  $p_i$ and $p_j$ determine the probability that individual $i$ defeats individual $j$ in a single pairwise comparison: 
\begin{eqnarray}
Pr(i \: \textrm{defeats} \: j) = \frac{p_i}{p_i+p_j}
\label{eq:BTlikelihood}
\end{eqnarray}
with
\begin{eqnarray}
\sum_{i=1}^{20}{p_i} = 1.
\end{eqnarray}
A maximum likelihood estimate for the twenty $p_i$ was extracted from the survey data by an iterative procedure \cite{RefWorks:919} \textcolor{black}{that rapidly converges to produce the optimal $p_i$; that is, it find the $p_i$ for which the dataset is most probable.  In addition, to test how robustly our particular dataset determined those  $p_i$, we performed $\sim 2000$ bootstrap random samplings of the maximum likelihood estimation. The bootstrap method yields an estimate for confidence levels in model parameters, reflective of the size and the internal self-consistency of the dataset. The uncertainties $\delta p$ in the reported $p_i$ are the $ 1 \: \sigma$ deviations obtained from the bootstrap test.}

\subsection*{Internet-based metrics of fame}		

\textcolor{black}{We then sought to test how other plausible metrics for fame correlate with the rankings obtained from the survey. Some possible metrics for fame are problematic as they are not universally applicable or cannot readily be measured or estimated for non-celebrities or for living individuals, or they are weighted toward people in certain professions, or they are controlled by gatekeepers. These include an individual's wealth, the length of his/her obituary or \emph{Who's Who} entry, numbers of Twitter followers, etc. Instead we sought to evaluate metrics that were (1) available for a wide range of individuals of diverse profession and varying fame, (2) reflective of the opinion of a large population or audience, rather than the judgment of curators or gatekeepers, (3) regularly updated, and (4) readily accessible through the internet. Based on these criteria, we selected the following plausible, internet-derived metrics of fame (\nameref{S2_Spreadsheet}) and evaluated them for the individuals on the NYT, NBC and Wiki lists:}
\begin{itemize}
\item \emph{GH} - the total current Google hits returned for the individual; 
\item \emph{GN} - the total current Google news items citing that individual;
\item \emph{WE} - the total edits to date of the individual's Wikipedia page; 
\item \emph{WV} - the total Wikipedia page views to date.  
\end{itemize}  
Most of the internet metrics for the names on the NYT, NBC and Wiki lists were assessed on March 8, 2017.  The data for the NYT 2 list was assessed on July 12, 2017, and the Wikipedia page views (\emph{WV}) were recorded on June 29, 2017.

Total current Google hits (\emph{GH}) was obtained by searching an individual's name in Google and counting the number of links returned.  Total current Google news items (\emph{GN}) was obtained by searching an individual by name and profession in Google News and counting the number of links returned.  For \emph{GN} searches where the individual could be identified with more than one profession, the search that returned the most links was used.	Total current Wikipedia page edits (\emph{WE}) were obtained from an individual's Wikipedia page through the ``History'' feature. 

To evaluate the temporal stability of these metrics we also retrieved time series data:  $WE_t$ is the month-to-month time series of Wikipedia page edits, obtained from the Revision History Statistics of the Wikipedia page;  $GS_t$ is the history of monthly Google searches, obtained from Google Trends; $WV_t$ is the daily history of Wikipedia page views, using Wikipedia PageViews Analysis.

\subsection*{Power law analysis}
We used a maximum likelihood method to assess whether metric \emph{WE} exhibits a power law distribution in the three lists studied \cite{RefWorks:913}. If $x$ is a discrete random variable whose probability distribution $p(x)$ for $x \ge x_{min}$ is a power law (Eq.~\ref{eq:powerlaw}) then the value of $\alpha$ for a dataset ${x_i}$ ($i = 1 \ldots n$) is estimated by maximizing the logarithmic likelihood of the data  
\begin{eqnarray}
L =  n \log(C) - \alpha \sum_{x_i \ge x_{min}}{\log(x_i)}
\end{eqnarray}
As the data  ${x_i}$ will not obey the power law below $x_{min}$, an estimate for $x_{min}$ is also needed.  For each name list we generated these estimates by minimizing the Kolmogorov-Smirnov distance between the cumulative distribution function (CDF) of the \emph{WE} data and that of a perfect power law \cite{RefWorks:913}. 

\subsection*{Data sharing}
Datasets are provided as supplemental information in \nameref{S1_Spreadsheet} and \nameref{S2_Spreadsheet}.

\subsection*{Use of human subjects}

Undergraduate students completed the fame $p_i$ survey under protocol 
IRB201700835, which was approved as exempt by the University of Florida Institutional Review Board (Behavior/Nonmedical, IRB-02). \textcolor{black}{Volunteer subjects were recruited in mid-June 2017, from public areas of the University of Florida campus.} Each subject read an informed consent document and provided oral consent for participation. The written consent requirement was waived owing to the minimal risk and the fact that no sensitive or identifying information was collected from the subjects.

\section*{Results and Discussion}


\subsection*{Survey results }

\begin{figure}[!h]
\includegraphics[width=.9\textwidth]{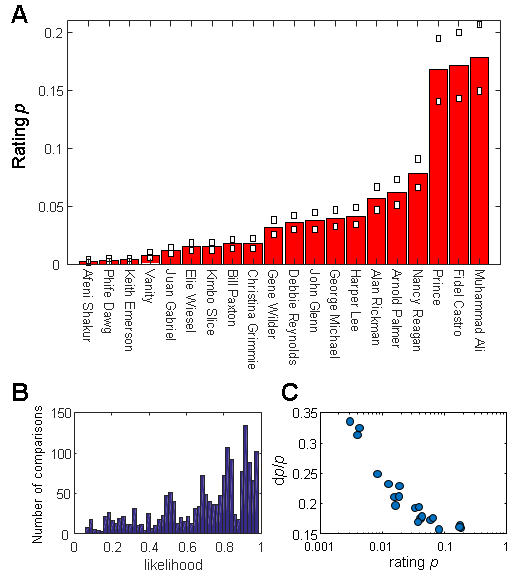}
\caption{\textbf{Human subject ratings of fame.} (A) Ratings $p$ of the fame of the individuals in Table~\ref{table1}.  Ratings were obtained by applying a Bradley-Terry ranking model (Eq.~\ref{eq:BTlikelihood}) to data from a survey in which subjects made pairwise comparisons of individuals listed in the table (\emph{Methods}). White squares show uncertainty estimates $\delta p$ ($\pm1 \: \sigma$) estimated from bootstrap analysis of survey data. (B) Histogram of likelihoods of the 1679 pairwise comparisons made by survey subjects, based on the $p$ values of (A) and Eq.~\ref{eq:BTlikelihood}. Given the $p$, 78\% of the comparisons have likelihood $  > 0.5$. (C) Scatterplot showing relative uncertainties $\delta p/p$ versus $p$ for the individuals in (A). }
\label{fig1}
\end{figure}

\textcolor{black}{We used a Bradley-Terry model \cite{RefWorks:912} to extract from the survey data a quantitative score of renown or fame for each of the  individuals in Table~\ref{table1}. These scores are the $p$ values that are shown in Table~\ref{table1} and Fig.~\ref{fig1}. The $p$ of each individual is a measure of his/her degree of renown, as derived from the set of pairwise comparisons or preferences reported by all the survey subjects. The maximum likelihood method described in \emph{Methods} identifies the unique, self-consistent set of $p$ values for which the entire survey dataset of 1679 subject preferences is most probable, based on Eq.~\ref{eq:BTlikelihood}. The obtained $p$  range over almost two orders of magnitude, from a maximum of  $0.18 \pm 0.03$ to a minimum of  $0.0029 \pm 0.0009$, indicating that the fame of the different individuals spans almost two decades, at least from the perspective of the subject population. }

\textcolor{black}{To assess whether (a) the number of survey subjects and (b) the number of preferences reported by those subjects were both sufficiently large, we examined the robustness of the $p$ values in Table~\ref{table1} using two different statistical tests.  First, as described in \textit{Methods}, we applied a bootstrap random sampling to evaluate the confidence intervals in the $p$ values, given our dataset.  The bootstrap method is a model-free approach that takes account of the size of the dataset as well as any lack of knowledge of the true or theoretical distribution of the model parameters. As shown in Fig~\ref{fig1}, the uncertainties $\delta p$ determined from the bootstrap correspond to relative uncertainties $\delta p /p$ of 10-30\%. As the less famous names in Table~\ref{table1} more frequently drew a ``no preference'' response from survey subjects, such names occur less often in the dataset; accordingly the bootstrap analysis finds a larger relative uncertainty $\delta p/p$ for these names. The relative uncertainty increases about two-fold from the best known ($\delta p /p \simeq 16 \%$ for $p \simeq 0.2$) to the least known ($\delta p/p \simeq 30\%$ for $p  \simeq 3\times 10^{-3}$) individuals. Nevertheless these relative uncertainties are still substantially smaller than most of the name-to-name differences in $p$ values.  This analysis shows that the survey dataset contains sufficient, self-consistent preference data to establish a robust ranking. }

\textcolor{black}{As a second statistical test of our survey sample and the model obtained from it, we also compared the relative likelihood of our findings (Fig~\ref{fig1}) to that of a null model for the same dataset. If for example the survey subjects are too few or are incompetent to rank the names usefully, then the relevant null model is one where the survey dataset contains too little information to support a significant ranking. In this null model all the names in Table~\ref{table1} have equal $p$ values, and either subject preference is equally likely for any name pair \cite{RefWorks:912}. Comparing the likelihood $L_{model}$ of our dataset under our model to its likelihood $L_{null}$ under the null model we find a very high log likelihood ratio $\log (L_{model}/ L_{null}) = 372$. That is, our $p$ values provide a $10^{161}$-fold better explanation of our dataset than does the null model. This is illustrated by Fig.~\ref{fig1}, where roughly 78\% of the 1679 preferences obtained from the survey have likelihood greater than 0.5, meaning that they are more likely than not, given our $p$ values and Eq.~\ref{eq:BTlikelihood}. The high likelihood of the dataset, given the $p$ values, demonstrates that the survey generated statistically significant, self-consistent information about the relative fame of the 20 individuals.}


Table~\ref{table1} shows the fame measures $p$ and $\delta p$, \emph{WE}, \emph{GN}, and \emph{GH} for the twenty individuals who died in 2016, identified by name and dates of birth (DOB) and death (DOD). It also shows $dWE/dt$, the average \emph{WE} added per month from the creation of the page through June 2017, and $dWV/dt$, the average  \emph{WV} per day from July 1 2015 to June 29 2017.

\subsection*{Testing correlation of fame metrics with $p$}

\begin{figure}[!h]
\includegraphics[width=.9\textwidth]{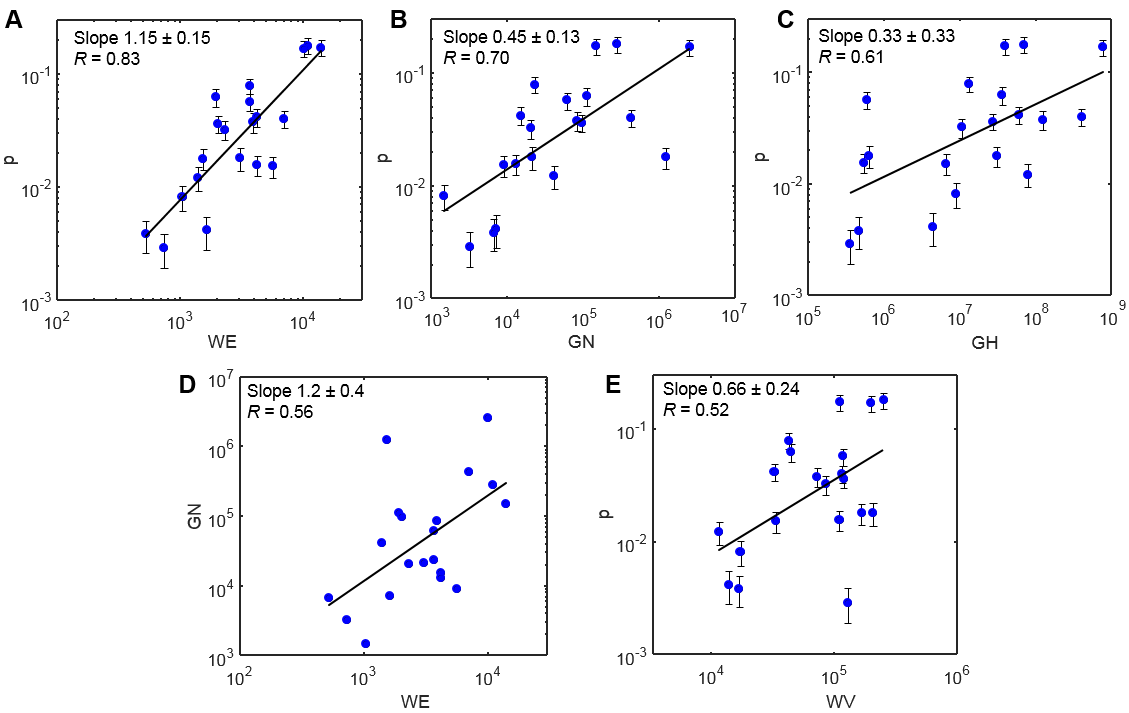}
\caption{\textbf{Correlations between quantitative measures of fame}.  Correlations between different quantitative measures of fame for the 20 individuals in Table \ref{table1}.  The survey metric $p$ is compared with internet metrics \emph{WE} (A), \emph{GN} (B), \emph{GH} (C), and \emph{WV} (E).  \emph{GN} and \emph{WE} are compared in (D).    In each double logarithmic plot the slope of the best (least squares) fit line is indicated. Panels are presented in order of decreasing Pearson correlation \emph{R} on the double log plot. Error bars for $p$ are obtained from bootstrap analysis. }
\label{Figure_Correlations}
\end{figure}

Therefore, using the individuals in Table~\ref{table1} as a test population, we evaluated several plausible internet-derived metrics of fame by testing their correlation with the $p$ values.  Like the $p$ values, the metrics \emph{GN}, \emph{GH}, \emph{WE} and \emph{WV} all range over several orders of magnitude. Fig.~\ref{Figure_Correlations} compares them against $p_i$ in order of decreasing Pearson correlation coefficient  $R$  of a double logarithmic plot.   \emph{WE} and \emph{GN} both show strong correlations, $R = 0.83$ and 0.70 respectively, with $p$.  \emph{WE} also has a nearly linear relationship with $p$, consistent with $p \propto \textrm{WE}^{1.2 \pm 0.2}$.  


Our data indicate that Google hits \emph{GH} and Wikipedia page views \emph{WV} are less reliable metrics of fame. \emph{WV} has a moderate correlation with $p$, giving $R = 0.52$. \emph{GH} has a weak correlation with $p$ ($R \simeq 0.6$), leading to a log-log slope $0.33 \pm 0.33$ that is consistent with zero. Although \emph{GH} has been regarded as an obvious metric of fame, the expansion of the internet may have made it less useful for distinguishing non-celebrities: A very high \emph{GH} ($\sim 10^6-10^7$) does correlate with celebrity status, but many common names have $GH \sim 10^5-10^6$ or higher, and therefore do not distinguish greater and lesser fame. Other flaws in \emph{GH} have also been noted \cite{RefWorks:867}.

\textcolor{black}{Overall we find that \emph{WE} and to a lesser extent \emph{GN} correlate sufficiently well with $p$ values and with each other that they may serve as useful quantitative measures of fame. However, we regard \emph{GN} and \emph{WE} only as metrics of the current fame of individuals, measured at a particular instant. Although it is likely that some individuals were more famous in the past than at their death, we do not attempt to construct a model to estimate their fame at its peak or to correct for any decline. In addition these internet-based metrics are probably not useful for comparing the fame of individuals who died at different times in the past. Clearly an individual who died prior to the Wikipedia launch in 2001 is less likely to acquire \emph{WE} than is a living person of otherwise comparable renown. Therefore in what follows we make no attempt to compare the fame of the recently deceased to that of individuals who died in earlier years. }

\subsection*{Time dependence} 

\begin{figure}[!h]
\includegraphics[width=.9\textwidth]{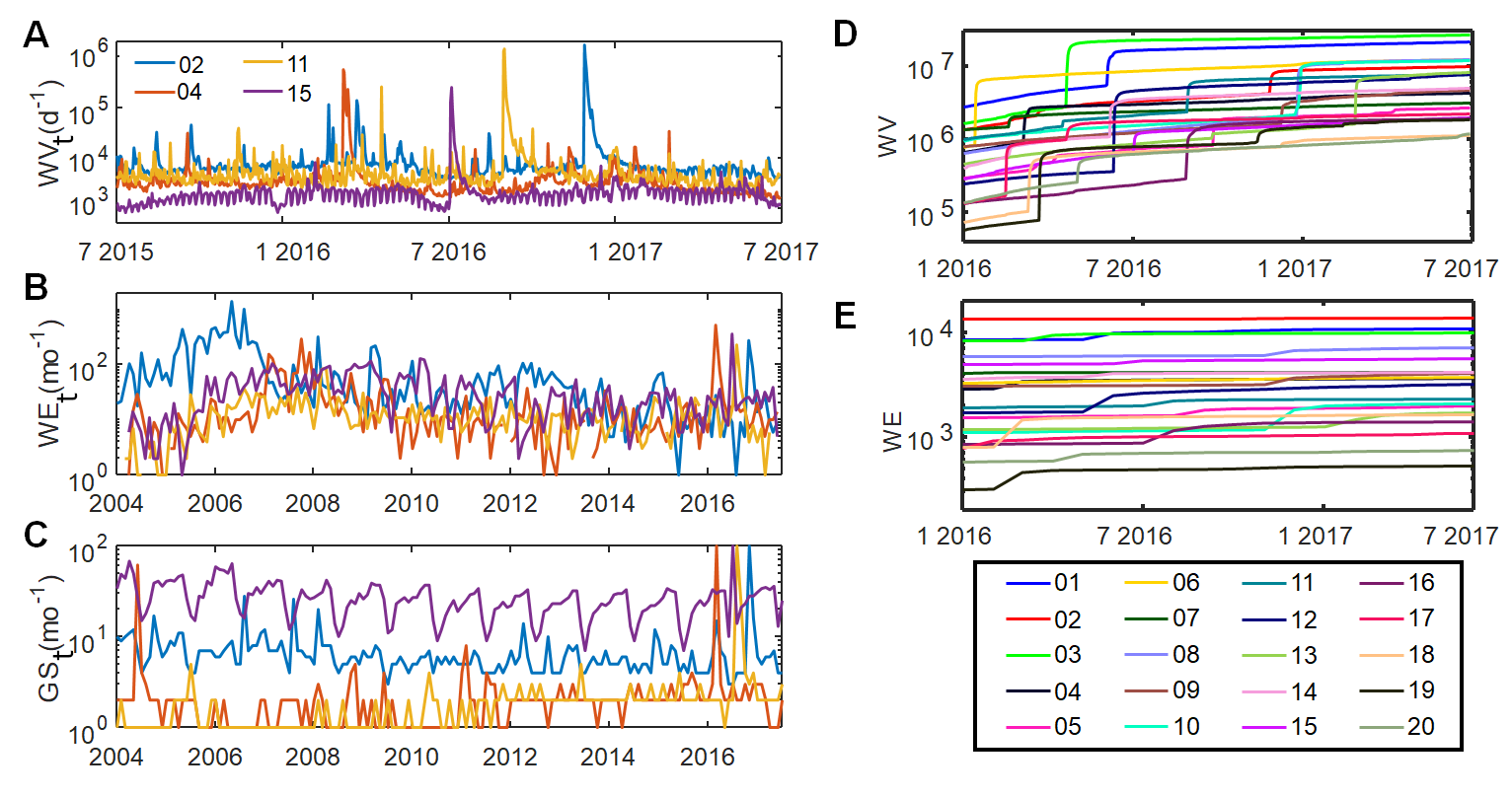}
\caption{\textbf{Temporal fluctuation and cumulative values of internet metrics of fame}. For four individuals in Table \ref{table1}, the figures at left show (A) the variation in daily Wikipedia page views ($WV_t$) over a two year interval, (B) monthly Wikipedia page edits ($WE_t$) over 13 years, (C) monthly Google searches, ($GS_t$), over 13 years.  $GS_t$ data for each individual are normalized to a maximum value of 100.  For all 20 individuals in Table~\ref{table1}, the figures at right show (D) cumulative \emph{WV} and (E) cumulative \emph{WE}.  The legends identify the individuals by their ID in Table \ref{table1}.  }
\label{Figure_TimeDependences}
\end{figure}

As an alternative to cumulative quantities such as total Wikipedia page edits we also evaluated some continuously varying indicators of fame, such as the monthly Wikipedia page edits $WE_t$.  Fig.~\ref{Figure_TimeDependences} shows several years of data for $WE_t$, $WV_t$ and $GS_t$ (denoting time series for \emph{WE}, \emph{WV} and Google searches/\textcolor{black}{Google Trends}, respectively) for four individuals in Table~\ref{table1}. Although $WE_t$ is noisy its behavior is generally stable and similar for all four individuals, with a dynamic range of about 10-100. By contrast $WV_t$ is subject to abrupt spikes associated with news events.  For some individuals, especially ID 15 and ID 02 in Table~\ref{table1},  $GS_t$ and $WV_t$ show strong weekly or annual periodicity, presumably due to regular cycles of student academic assignments.  The instability of $GS_t$ and $WV_t$ argues against the use of short-term snapshots of \emph{GS} or \emph{WV} as quantitative metrics of fame.


Fig.~\ref{Figure_TimeDependences} also shows the accumulation of total \emph{WE} and \emph{WV} over a 1.5 y interval, for the individuals in Table~\ref{table1}.  While an individual's \emph{WV} may jump discontinuously when the individual's death is reported,  \emph{WE} generally changes more slowly and its relative ordering is largely stable over time.   These data suggest that while both \emph{WV} and \emph{WE} inevitably increase over time, a rank ordering of individuals by \emph{WE} changes slowly, as required for a useful quantitative measure of fame.

\begin{figure}[!h]
\includegraphics[width=.9\textwidth]{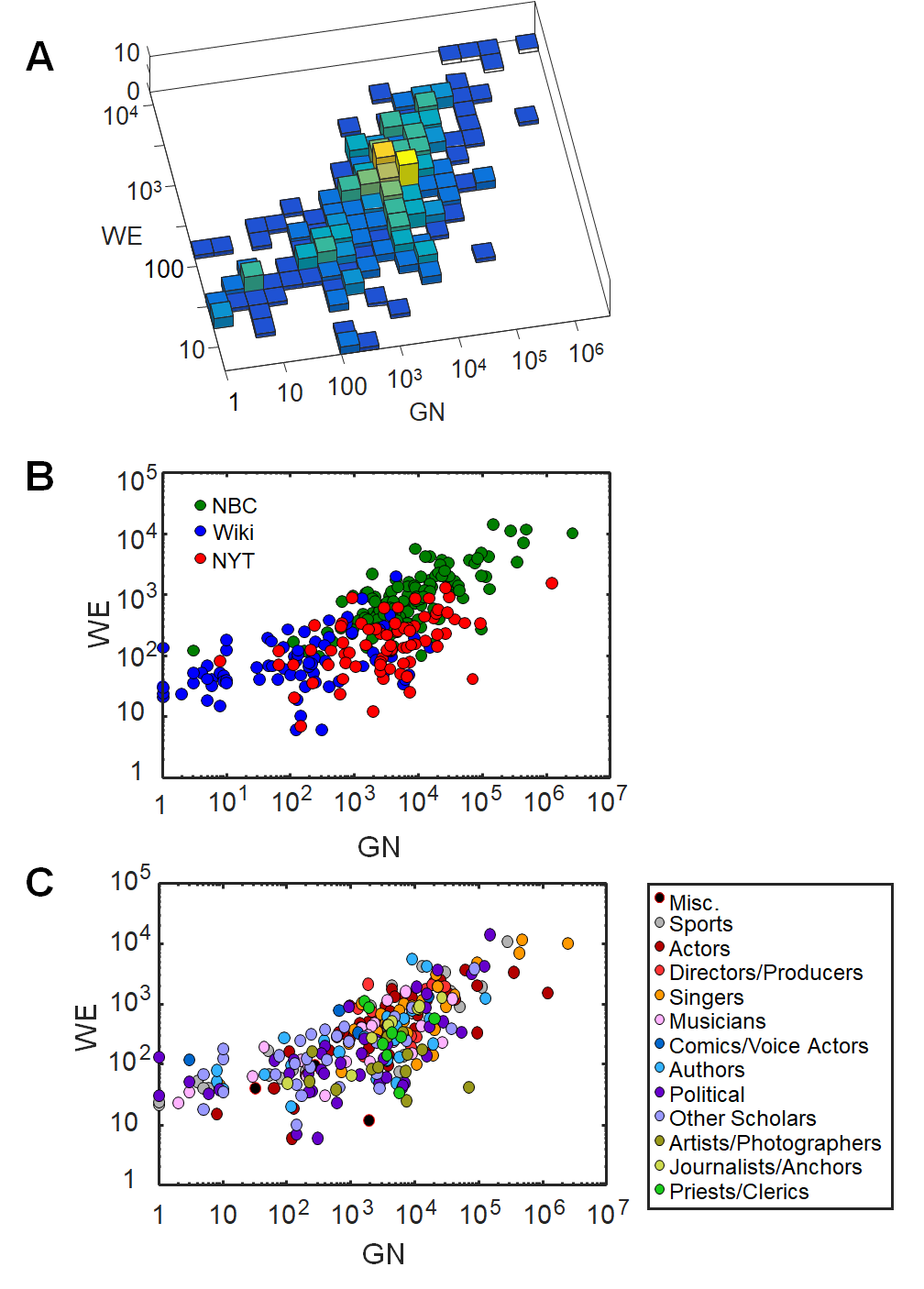}
\caption{\textbf{The correlation between \emph{GN} and \emph{WE}.} The correlation between \emph{GN} and \emph{WE}, as measures of fame for the individuals in the three datasets (NBC, Wiki and NYT 1), is shown in three double-logarithmic plots.  (A) The logarithms of \emph{GN} and \emph{WE} are reasonably well correlated, with a Pearson $R = 0.71$.  (B) \emph{WE} and \emph{GN} show similar correlation in the three datasets, which sample different ranges of \emph{GN}; (C)   \emph{WE} and \emph{GN} are shown with a color code that indicates the profession of the individuals.    }
\label{WEGNScatterplot}
\end{figure}

As \emph{WE} and \emph{GN} both correlate reasonably well with $p$, we expect them to correlate with each other. The scatterplot of Fig.~\ref{WEGNScatterplot} shows that \emph{GN} and \emph{WE} correlate similarly in all three datasets (279 individuals). The data fall roughly along a curve that is more nearly quadratic ($GN \propto WE^2$) than linear, so that while \emph{GN} spans more than six decades, \emph{WE} spans about 3-4 decades. One possible interpretation of the nonlinear relationship is that \emph{WE}, which is subject to the practical aspects of web editing, is unlikely to be smaller than about 10, and therefore has a floor value even though \emph{GN} does not. Another possibility is that \emph{GN} is more sensitive to celebrity (as defined herein) whereas \emph{WE} is a better measure of fame, so that \emph{GN} emphasizes more famous individuals at the expense of the less famous.  


Fig.~\ref{WEGNScatterplot}C illustrates the relation between $GN$ and $WE$ for individuals of different professions.  This and other analysis we conducted show no evidence that the correlation between \emph{WE} and \emph{GN} depends significantly on profession.   

\subsection*{The probability distribution for fame}

\begin{figure}[!h]
\includegraphics[width=.9\textwidth]{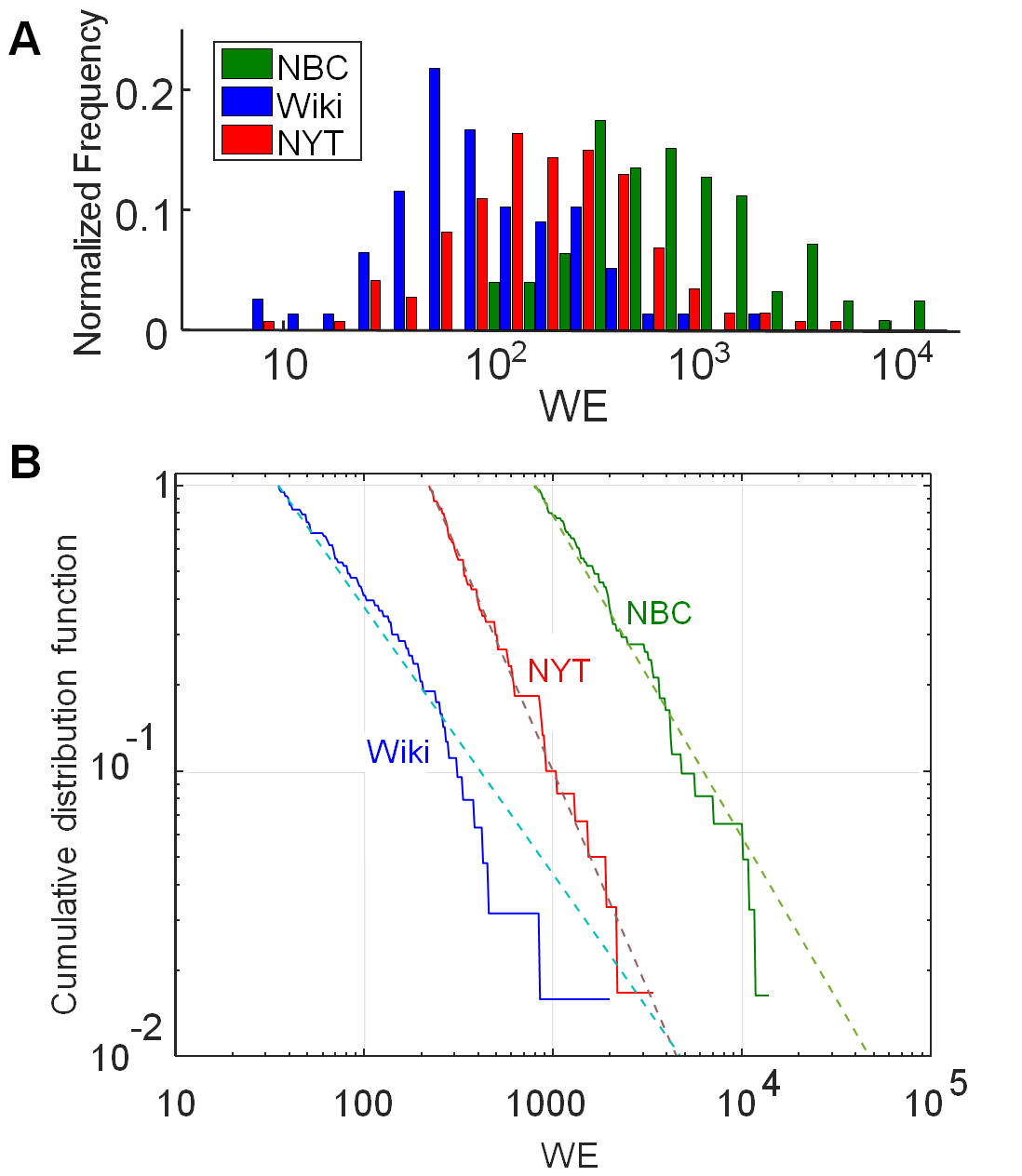}
\caption{\textbf{Probability distributions for fame.}	Probability distributions for fame, as measured by \emph{WE}, for individuals in the NBC, Wiki, and NYT (1 \& 2) datasets.  (A) Histogram showing, for each dataset, the distribution of reported deaths with respect to \emph{WE}. (B) Cumulative distribution functions (CDF) with respect to \emph{WE} are shown as the solid curves.   CDF are evaluated using cutoff \emph{WE} values of 35 (Wiki), 220 (NYT) and 800 (NBC).  For each dataset, the optimal value of the power law exponent and cutoff ($\alpha$  and $x_{min}$ respectively in Eq.~\ref{eq:powerlaw} ) were evaluated as described in \emph{Methods} \cite{RefWorks:913}, with results shown in Table~\ref{Table:PowerlawParameters}. The dashed lines show the CDF for true power law distributions that have the same $\alpha$ and $x_{min}$ as the data.  }
\label{Figure_PowerLaw}
\end{figure}

The survey-generated $p_i$  provide an intuitive and quantitative metric of fame.  However the printed survey is impractical for evaluating the fame of larger numbers of individuals. Therefore we use \emph{WE}, which appears to be a satisfactory alternative metric of fame, to investigate the statistical distribution of fame. Fig.~\ref{Figure_PowerLaw}A shows histograms of \emph{WE} for individuals in the NYT, NBC and Wiki datasets. In each case the \emph{WE} distribution is broad, spanning at least two decades, with a tail extending to very large \emph{WE}. The tails raise the question of whether fame, like many other quantities in the social and natural sciences, obeys a power law distribution. Phenomena such as forest fires, earthquakes, and the sizes of cities, which lack an intrinsic size scale \cite{RefWorks:907}, often obey a power law: The probability that an event has magnitude $x$ is given by
\begin{eqnarray}
p(x) = \frac{C}{x^\alpha}
\label{eq:powerlaw}
\end{eqnarray}
\begin{eqnarray}
\sum_{x=x_{min}}^{\infty}{p(x)} =1
\end{eqnarray}
for $x > x_{min}$. Here $x_{min}$ is a cutoff, $C$ is a normalization constant, and we have taken $x$ as a discrete variable (like \emph{WE}).


 \textcolor{black}{We tested our data for power law behavior by finding the power law model that best fit the cumulative distribution function (CDF) of each dataset. (The CDF is the function $F(x)$ that gives the probability that any one measurement $X$ exceeds $x$.)  For each dataset we assumed that the \emph{WE} data obey Eq~\ref{eq:powerlaw} above a cutoff value of \emph{WE} (\emph{Methods}). As shown in Table~\ref{Table:PowerlawParameters}, all three datasets give comparable values, $\alpha \simeq 1.9 - 2.6$, although with very different cutoff ($x_{min}$) values, indicative of the different selectivity of the three data sources. Fig.~\ref{Figure_PowerLaw} illustrates the agreement by showing the cumulative distribution function (CDF) for \emph{WE} and that of the corresponding, maximum-likelihood power law. Many apparent power laws are only approximate and do not withstand close statistical scrutiny \cite{RefWorks:913}.  Although the size of NYT and NBC datasets is insufficient to establish whether the power law is superior to other models for the distribution, these particular datasets do appear to show good agreement with the power law model.}  We note that our $\alpha$ values are consistent with the  $\alpha = 1.9-2.1$ that was estimated by a different method in a study of the fame of WWI flying aces \cite{RefWorks:859}. 

\renewcommand{\arraystretch}{1.4}

\begin{table}[!ht]
	\begin{adjustwidth}{-2.0in}{0in}
	\centering
	\caption{Results of fitting Eq.~\ref{eq:powerlaw} and Eq.~\ref{eq:GR} to \emph{WE} data }
	\label{Table:PowerlawParameters}
	\begin{tabular}{lllllll}
	& \cellcolor[HTML]{C0C0C0}{\color[HTML]{333333} \textbf{\boldmath{Wiki}}} & \cellcolor[HTML]{C0C0C0}{\color[HTML]{333333} \textbf{\boldmath{NBC}}} & \cellcolor[HTML]{C0C0C0}{\color[HTML]{333333} \textbf{\boldmath{NYT}}}  \\ \cline{2-7} 
	\multicolumn{1}{l|}{\cellcolor[HTML]{C0C0C0}{\color[HTML]{333333} \textbf{\boldmath{$x_{min}$}}}}    & \multicolumn{1}{l|}{20-70}                                          & \multicolumn{1}{l|}{700-900}                                          & \multicolumn{1}{l|}{220-240}                                          \\ \cline{2-7} 
	\multicolumn{1}{l|}{\cellcolor[HTML]{C0C0C0}{\color[HTML]{333333} \textbf{\boldmath{$\alpha $}}}} & \multicolumn{1}{l|}{$1.9 \pm 0.1$}                                    & \multicolumn{1}{l|}{$2.1 \pm 0.1$}                                     & \multicolumn{1}{l|}{$2.6 \pm 0.2$}                                    \\ \cline{2-7} 
		\multicolumn{1}{l|}{\cellcolor[HTML]{C0C0C0}{\color[HTML]{333333} \textbf{\boldmath{$\nu$}}}}    & \multicolumn{1}{l|}{$1.7 \pm .2$}                                          & \multicolumn{1}{l|}{$1.5\pm 0.1$}                                          & \multicolumn{1}{l|}{$1.7 \pm 0.1$}                                          \\ \cline{2-7} 
				\multicolumn{1}{l|}{\cellcolor[HTML]{C0C0C0}{\color[HTML]{333333} \textbf{\boldmath{$a$}}}}    & \multicolumn{1}{l|}{$\simeq 0.00013$ y}                                          & \multicolumn{1}{l|}{$\simeq 0.0079$ y}                                          & \multicolumn{1}{l|}{$\simeq 0.0011$ y}                                          \\ \cline{2-7} 
								\multicolumn{1}{l|}{\cellcolor[HTML]{C0C0C0}{\color[HTML]{333333} \textbf{\boldmath{$b$}}}}    & \multicolumn{1}{l|}{$\simeq 8\times10^6$}                                          & \multicolumn{1}{l|}{$\simeq 3\times10^6$}                                          & \multicolumn{1}{l|}{$\simeq 4\times10^6$}                                          \\ \cline{2-7} 
\end{tabular}
\end{adjustwidth}
\end{table} 

\begin{figure}
\includegraphics[width=.9\textwidth]{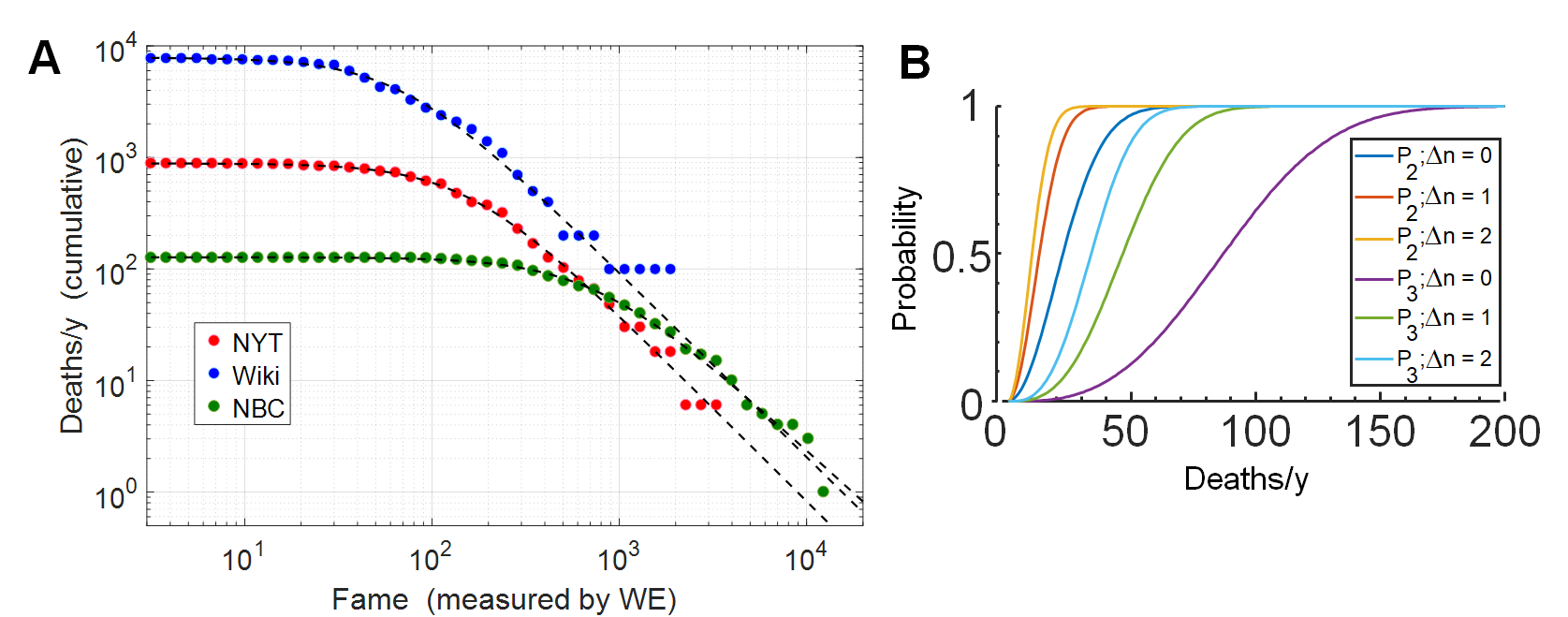}
\caption{\textbf{Cumulative frequency of deaths versus fame.}	 (A) Cumulative frequency (Gutenberg-Richter) plot of reported deaths versus fame. For each degree of fame (as measured by \emph{WE}), the plot shows the annualized rate at which deaths of equal or greater fame occur. Green, red, and blue symbols indicate the NBC, NYT (1 \& 2) and Wiki datasets respectively.     Dashed curves show least squares fit to Gutenberg-Richter expression, Eq.~\ref{eq:GR}, giving $\nu = 1.5 -  1.7$ (Table~\ref{Table:PowerlawParameters}).  (B) Date overlap (birthday-problem) probability versus death rate. Each curve gives the probability $P$ that, if the indicated number of deaths occur at random during one year, there will be at least one instance where two  ($P_2$) or three ($P_3$) of those deaths are separated by only $\Delta n$ days.} 
\label{GRPlot}
\end{figure}

\textcolor{black}{While the CDF gives the probability that any particular event exceeds a certain magnitude, it is often more useful to know the absolute frequency of events of a certain magnitude.  For example in the case of earthquakes it is helpful to know how many events exceeding a given threshold occur each year. A cumulative frequency plot shows the frequency $f(x)$ of events that have magnitude greater than or equal to $x$.  For many natural power law phenomena the cumulative frequency plot obeys the empirical Gutenberg Richter law  \cite{RefWorks:907}}

\begin{eqnarray}
f(x) = \frac{1}{a + x^{\nu}/b}
\label{eq:GR}
\end{eqnarray}
Here $a$ and $b$ set the overall scale of the frequency and define a threshold sensitivity in the dataset (similar to $x_{min}$ above), while $\nu$ reflects the power law distribution of the underlying events. Fig.~\ref{GRPlot} shows the cumulative frequency plots for all three datasets, using \emph{WE} as a fame measure and scaling the event numbers in the data up to equivalent annual rates. The parameters $a$ and $b$, shown in Table~\ref{Table:PowerlawParameters}, are highly variable as the different media sources apply different selectivity criteria in reporting obituaries. However as \emph{WE} increases, all three datasets tend toward Gutenberg-Richter behavior with $\nu \simeq 1.5-1.7$ (by least squares fit).


It is interesting that all three curves indicate that $\sim 30-100$ individuals with fame $WE \ge 10^3$ die each year. Such information can yield quantitative insight into questions that arise perennially about the frequency of famous deaths \cite{RefWorks:954} \cite{RefWorks:955} \cite{RefWorks:977}. For example, the so-called celebrity rule of three is a folk belief that deaths of celebrities occur in clusters, especially groups of three, spread over a few days \cite{RefWorks:955}.   A frequently mentioned example is the three days in June 23-25, 2009 during which the television personality Ed McMahon, the musician Michael Jackson and the actress Farrah Fawcett all died. Another striking example is the deaths of C.S. Lewis, Aldous Huxley, and John F. Kennedy, on November 22, 1963. 

The probability of such coincidences can be estimated from the data in Fig.~\ref{GRPlot} by defining a threshold for fame, referring to the Gutenberg-Richter plot, and then applying the familiar birthday problem in statistics:  If the deaths of $N$ individuals are randomly distributed throughout the year, then if $N \ge 23$ the probability exceeds 50\% that at least two deaths will occur on the same day. If the threshold for fame is $WE \ge 1000$, the expected 30-100 deaths per year ensures that two famous deaths will occasionally coincide. 

The common statement of the celebrity rule of three does not require the deaths to occur on precisely the same day.  If they may be separated by $\Delta n$ days, then as shown in Fig.~\ref{GRPlot} a 50\% probability of two occurring in coincidence requires only $N \ge 14$ (for $\Delta n = 1$) or $N \ge 11$ ($\Delta n = 2$) deaths per year. Among individuals of rather high fame, \emph{WE} $\ge 2000-3000$, at least one such coincidence appears likely each year. A three-person coincidence has 50\% probability when $N \ge 88$ ($\Delta n = 0$) or $N \ge 35$ ($\Delta n = 2$):  If $\Delta n \le 2$ is considered a coincidence, then at least one cluster of three deaths with $ WE > \sim 1000$ seems likely to occur in most years. \textcolor{black}{Therefore, although Fig.~\ref{GRPlot} summarizes only one year of deaths, the data are sufficient to demonstrate that the apparent clustering of famous deaths is not an entirely false perception; rather the clustering is a statistical consequence of the rather large number of famous deaths that occur each year.}    


\section*{Conclusion}
The application of statistical ideas has led productively to greater understanding of many aspects of human social dynamics, such as the evolution of opinion, cultural and linguistic behaviors \cite{RefWorks:910}. Our results demonstrate that quantitative measures can plausibly be applied to fame, providing insight into this important cultural phenomenon and allowing detailed statistical investigation. We note however that, as fame has economic value, it would be preferable to measure fame using tools that (unlike \emph{WE}) are difficult to manipulate. For example, instead of using paper surveys to score fame, one could implement a larger scale, social-media based, electronic version of the pairwise comparison method. This would greatly expand the survey base, allowing more accurate evaluation of the fame of greater numbers of individuals, and especially less renowned individuals.  Future investigators may wish to explore such approaches.

\section*{Supporting information}


\paragraph*{S1 Spreadsheet.}
\label{S1_Spreadsheet}  {\bf Survey Responses} The spreadsheet contains all individual responses to the fame survey. The first page contains the list of twenty deceased individuals of high renown, each identified by a number ID. Each of the 50 survey subjects was presented with 50 pairs of names from this list and asked to select the more familiar name in each pair.  

The second page contains the response data. The first column gives the subject number.  The second and third columns indicate the ID of the winner (more familiar) and loser (less familiar) in each pairwise comparison of names.  

Because ``No preference'' responses have been removed, fewer than fifty responses are recorded for some subjects.

\paragraph*{S2 Spreadsheet.}
\label{S2_Spreadsheet}  {\bf Fame metrics for individuals in the datasets} 
The spreadsheet contains fame metrics for (1)  the 20 individuals in Table~\ref{table1}; (2)  the 126 individuals in the NBC dataset; (3)  the 78 individuals in the Wiki dataset; (4) the 75 individuals in the NYT 1 dataset; (5) the 72 individuals in the NYT 2 dataset.  It also contains (6) the occupation codes used in the data tables (2)-(4) above.


%
%
%
\bibliography{references2} 

%
%
%
%

\end{document}